\documentclass[runningheads]{llncs}
\usepackage[T1]{fontenc}
\usepackage{graphicx}
\usepackage{booktabs} 
\usepackage{tabularx} 
\usepackage{paralist} 
\usepackage{multirow}
\usepackage{amsmath}
\usepackage{url}
\usepackage{color}

\newcommand{\Stanik}{Stanik dataset}
\newcommand{\Brunotte}{Brunotte dataset}

\newcommand{\Precision}{Prec.}
\newcommand{\Recall}{Rec.}
\newcommand{\FOne}{$F_1$}

\newcommand{\fix}[1]{\textcolor{black}{#1}}

\begin{document}
\title{Automatically Classifying Kano Model Factors in App Reviews}
%
%
\author{Michelle Binder\inst{1}\orcidID{0000-0002-8657-3963}\and
Annika Vogt\inst{1}\and
Adrian Bajraktari\inst{2}\orcidID{0000-0001-5997-7156}\and
Andreas Vogelsang\inst{2}\orcidID{0000-0003-1041-0815}}
\authorrunning{M. Binder et al.}
%
\institute{University of Cologne, Cologne, Germany \\
\email{\{mbinder1,avogt16\}@smail.uni-koeln.de}  \and
University of Cologne, Computer Science, Cologne, Germany\\
\email{\{bajraktari,vogelsang\}@cs.uni-koeln.de} 
}
\maketitle              
\begin{abstract}
\textbf{[Context and motivation]} 
Requirements assessment by means of the Kano model is common practice. As suggested by the original authors, these assessments are done by interviewing stakeholders and asking them about the level of satisfaction if a certain feature is well implemented and the level of dissatisfaction if a feature is not or not well implemented.   
\textbf{[Question/problem]} 
Assessments via interviews are time-consuming, expensive, and can only capture the opinion of a limited set of stakeholders. 
\textbf{[Principal ideas/results]} 
We investigate the possibility to extract Kano model factors (basic needs, performance factors, delighters, \fix{irrelevant}) from a large set of user feedback (i.e., app reviews). 
We implemented, trained, and tested several classifiers on a set of 2,592 reviews. In a 10-fold cross-validation, a BERT-based classifier performed best with an accuracy of 92.8\%. To assess the classifiers' generalization, we additionally tested them on another independent set of 1,622 app reviews. The accuracy of the best classifier dropped to 72.5\%. We also show that misclassifications correlate with human disagreement on the labels.
\textbf{[Contribution]} 
Our approach is a lightweight and automated alternative for identifying Kano model factors from a large set of user feedback. The limited accuracy of the approach is an inherent problem of missing information about the context in app reviews compared to comprehensive interviews, which also makes it hard for humans to extract the factors correctly.    

\end{abstract}
\section{Introduction}

Figuring out which features and \fix{related} requirements are important to stakeholders and, thus, should be implemented first or with special care is one of the core activities in Requirements Engineering (RE). There is a plethora of requirements prioritization techniques, in which usually costs and benefits are weighed up either by expert assessment or by stakeholder involvement~\cite{Bukhsh2020,Herrmann2008}.

One of the most well-known and applied techniques in requirements prioritization is the Kano model~\cite{kano1984}. It is based on the two-factor theory by Herzberg~et~al.~\cite{herzberg1993motivation}, which says that a factor that leads to satisfaction does not necessarily lead to dissatisfaction if absent and vice versa. The Kano model  categorizes product features into a set of five factors, which have different satisfaction--dissatisfaction profiles.

Several studies unanimously report that scalability is one of the major limitations of requirements prioritization techniques~\cite{Achimugu2014,Bukhsh2020,Hujainah2018}. This also holds for the assessment of \fix{product features} according to the Kano model. The categorization of \fix{product features} into the five Kano factors is done by interviewing or surveying stakeholders. This process is laborious and limited to the set of available stakeholders and a set of predefined requirements under investigation.

We investigate the possibility to \fix{identify} Kano model factors automatically from a large set of user feedback to increase scalability and broaden the focus to a large set of users and their specific feedback. 
We implemented, trained, and tested several classifiers to \fix{explore} their ability to identify Kano model factors in app reviews. 
\fix{The resulting categorization is, first of all, a categorization of user feedback, which may later be related to product features either manually or automatically~\cite{AlAmoudi2022,Guzman2014}.
}

\fix{We trained and evaluated our classifiers on two datasets}, which were collected by independent research groups and which we labeled manually. To do so, we created a labeling guideline and labeled a large part of the data independently by two labelers. We used random undersampling to create a balanced set of 2,592 app reviews from the larger of the two datasets. We used this dataset for training and testing the classifiers in terms of a 10-fold cross validation. We used the second dataset to test the classifiers on 1,622 unseen and independently collected app reviews. Finally, we compared misclassifications of the classifiers with the initial labels of the two human labelers and whether they agreed or disagreed initially. 
In this paper, we make the following contributions:
\begin{compactitem}
\item We find that Kano model factors can, to some degree, be automatically identified in app reviews. 
\item We find that our Kano model classifiers still lack sufficient generalization to other datasets.
\item We find that misclassifications, to some extent, correlate with human disagreement on the
labels.
\item We publish two datasets with 8,126 app reviews overall, with labels representing Kano model factors.
\end{compactitem}

Our approach is a lightweight and automated alternative for identifying Kano model factors from a large set of user feedback. The limited accuracy of the approach is an inherent problem of missing information about the context in app reviews compared to comprehensive interviews, which also makes it hard for humans to extract
the factors correctly.

\noindent\textbf{Availability of Data and Code:}
The datasets including the Kano model labels and the code for all classifiers and analysis procedures are publicly available.\footnote{\url{https://doi.org/10.6084/m9.figshare.21618858}}

\section{Background and Related Work}
In this section, we will introduce the Kano model and present related work in the field of app store analytics and natural language processing (NLP).

\subsection{Kano Model}
The Kano model~\cite{kano1984} describes the relationship between customer satisfaction and the implementation status of quality characteristics of a product. 
Kano distinguishes between five categories:
\begin{compactitem}
\item \textbf{Basic features} are perceived by users as intrinsic and ``normal'' for a product type. They only become aware of them when they are not available or not working (implicit expectations). Their presence does not lead to any satisfaction, while their absence leads to dissatisfaction.
\item \textbf{Performance features} lead to dissatisfaction if not or poorly implemented, while leading to satisfaction if fully implemented. 
\item \textbf{Delighters} are features the customer is not expecting. Their presence leads to satisfaction, while they do not lead to dissatisfaction if not implemented.
\item \textbf{Irrelevant features} lead neither to satisfaction nor to dissatisfaction.
\item \textbf{Rejection features} lead to dissatisfaction if they are implemented.
\end{compactitem}
In this paper, we ignore rejection features since their definition in the literature is ambiguous. The effect of their absence is described as causing dissatisfaction or as causing neither satisfaction nor dissatisfaction. Many sources only mention the first three factors of the Kano model.
\begin{figure}
    \centering
    \includegraphics[width=0.6\textwidth]{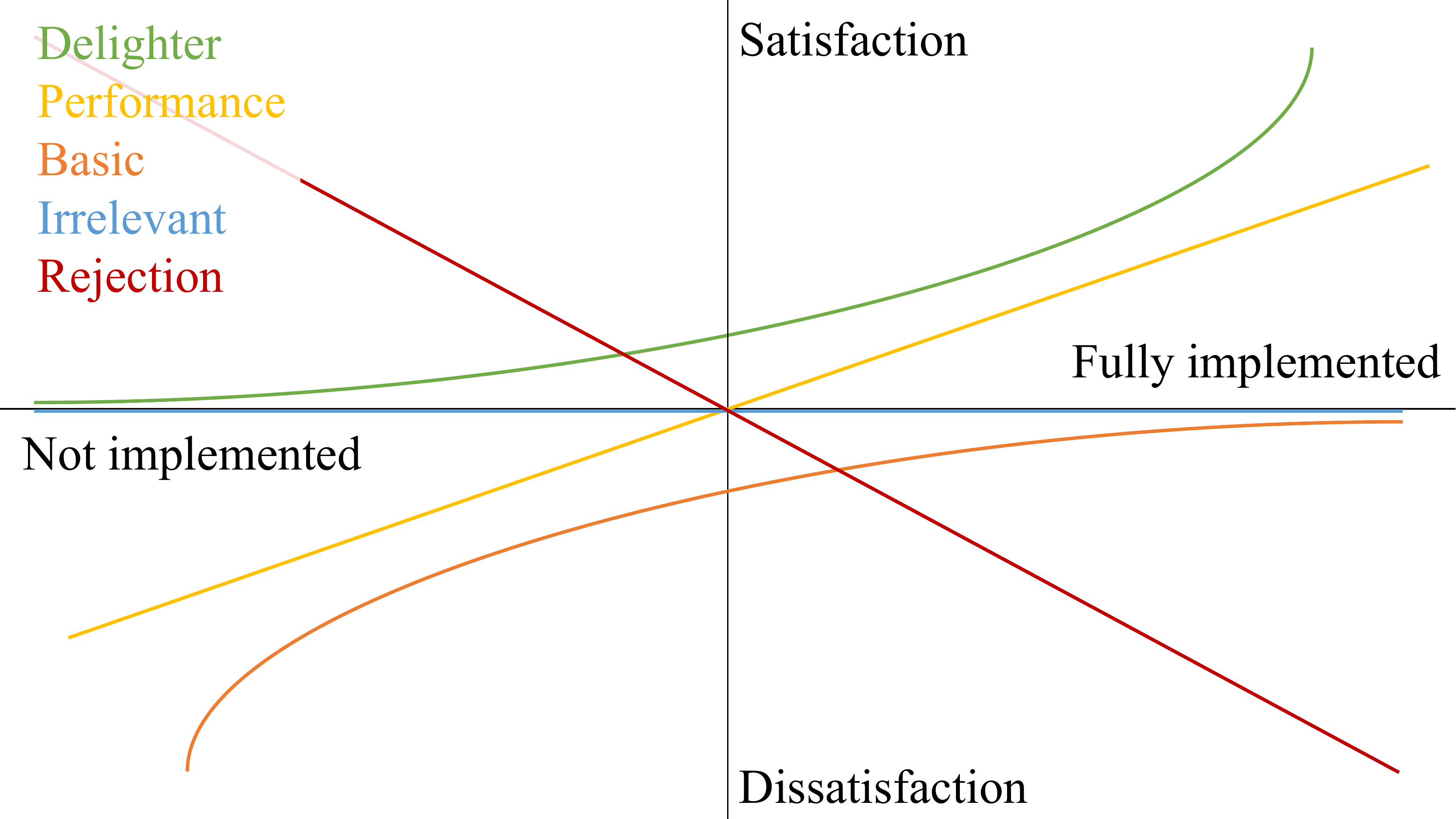}
    \caption{Visualization of the Kano model.}
    \label{fig:kano-model}
\end{figure}

Fig.~\ref{fig:kano-model} shows a common representation of the model including rejection features according to the former interpretation.
Features are rarely universal, i.e., what is a basic feature for one customer may be a performance feature for another, etc.
In addition, there is a temporal aspect to these categories. As time goes on, delighters become performance features and performance features become basic features.

In his original work, Kano proposed to conduct interviews with stakeholders to categorize features into these classes. However, due to the massive amount of users of apps and the amount of reviews in app stores, neither interviews nor manual categorization are feasible for app developers.

\subsection{Crowd-based RE and App Store Analytics}
User feedback is an important asset in the development of apps. A study~\cite{Maalej2016a} on the usage of analytics tools in app stores showed that tools only providing sales, download, and demographic data are not of high interest for developers. However, developers perceive tools that support app review analytics as helpful.
\fix{Wang et al.~\cite{Wang2019} did a systematic mapping study on crowd-sourced requirements engineering using user feedback. They found that, in many works, user feedback has been used in requirements elicitation and requirements analysis mainly, but also in requirements management.}
\fix{Wouters et al.~\cite{Wouters2022} created a method to integrate crowd-based requirements engineering into  development. The crowd is responsible for generating, voting, and discussing ideas, while the remaining activities are done by the development team.}
\fix{Lim et al.~\cite{Lim2021SLR} did a systematic literature review on data-driven requirements elicitation. They found that there are seven main sources of data used in the literature: online reviews, blogs, forums, software repositories, usage data, sensor readings and mailing list. Further, the main methods used were categorized as machine learning, rule-based classification, model-oriented, topic modeling and clustering.}

\fix{Reviews in app stores are a rich source of user feedback for crowd-based RE~\cite{Groen2017}}.
Pagano and Maalej~\cite{Pagano2013} conducted an empirical study on user feedback in app stores, showing that app stores can serve as communication channels between users and developers, allowing to continuously receive bug reports, feature requests, praise, etc. Developers can use reviews to understand new user needs since they provide more insight than plain statistics into how apps are actually used. They further find that tools should support automatic analyses of user feedback. 
\fix{There have been several approaches to automatic feature extraction, e.g., using sentiment analysis~\cite{Guzman2014} or a combination of NLP, metadata, text classification and sentiment analysis~\cite{Maalej2015}.}
Maalej et al.~\cite{Maalej2016b} found that review analytics tools are promising \fix{for review classification}, as a classification accuracy of 85\% to 92\% is possible. They conducted interviews with nine practitioners to evaluate their analytics tool. They do not often consider reviews, as the manual extraction of relevant information is too time consuming. Also, they usually gather input from multiple sources, e.g., emails, test groups, etc. However, they perceive user reviews as a promising source of information when assisted by tools to filter and categorize them automatically.



 

\fix{We found two papers, where the authors suggested automatic approaches to identify Kano model factors in user feedback.}
\fix{AlAmoudi et al.~\cite{AlAmoudi2022} analyzed app store reviews and categorized them according to the Kano model by using NLP techniques and clustering. They achieved high precision but low recall for basic features, high recall but low precision for delighter features and mid to low results on performance features.}
\fix{Lee et al.~\cite{Lee2022} used sentiment analysis to categorize hotel service ratings into Kano factors. They achieved rather low results and concluded that the reviews tend to be more about personal experience with a service, rather than an overall evaluation.}

\subsection{NLP and Machine Learning}

Natural language, due to its easy to write and comprehend nature, is the traditional way to document requirements. In the last decades, requirement engineers studied many aspects of NLP, ranging from modeling and abstracting key elements to automatic classification and clustering~\cite{Dalpiaz2018}. \fix{Two key challenges when using NLP for RE are availability of proper datasets and domain adaptation of models~\cite{Dalpiaz2018}.}
Applying NLP tools to RE task has developed from using
traditional machine learning techniques on hand-crafted features like bag of words~\cite{Maalej2015} to deep learning techniques where the input is encoded with word embeddings~\cite{Winkler2016}. Recently, transfer learning approaches that work on large pretrained language models (e.g., BERT) showed the best results for many RE tasks~\cite{Fischbach2023,Henao2021,Hey2020,Sainani2020}.




\section{Research Methodology}

\begin{figure}
    \centering
    \includegraphics[width=\textwidth]{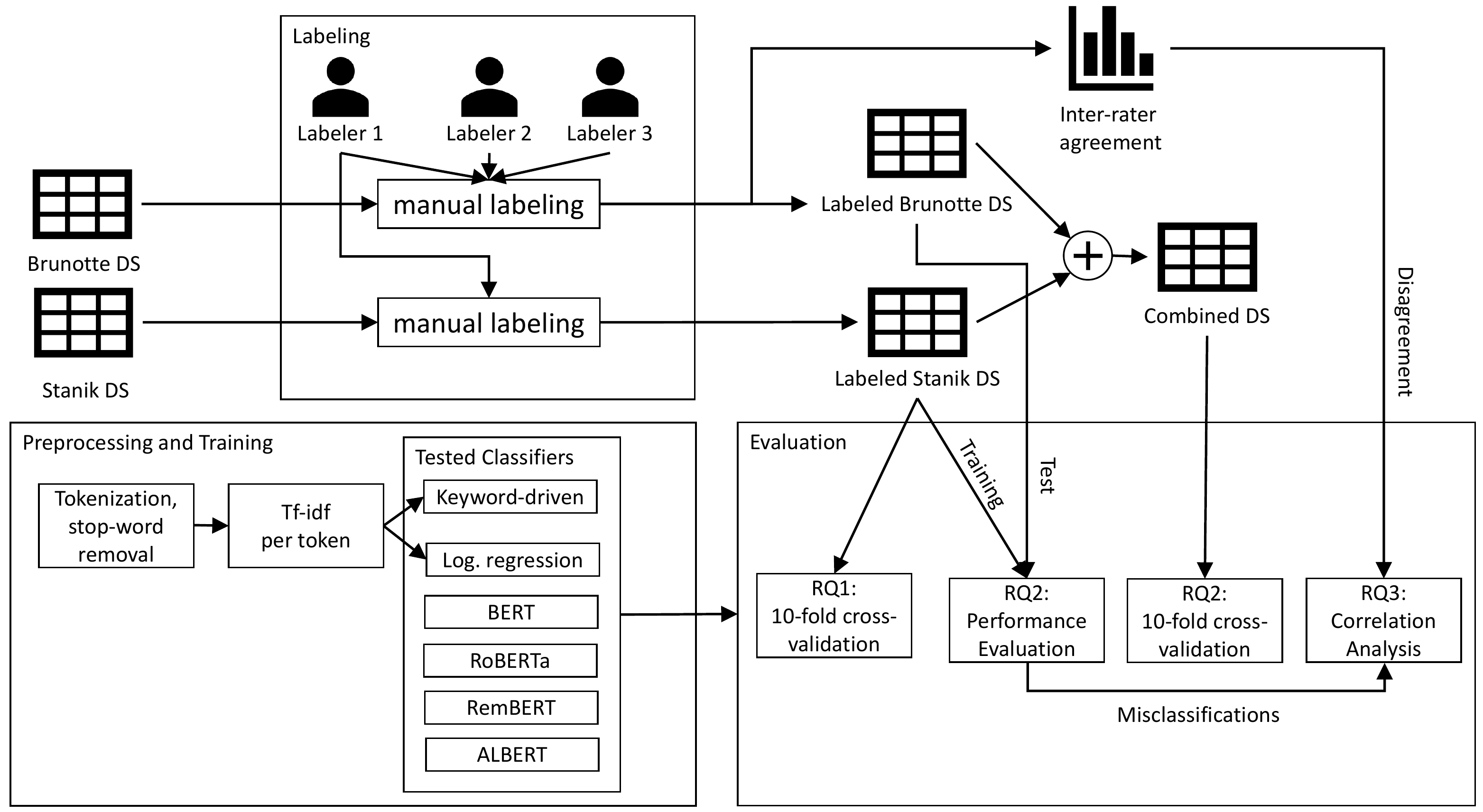}
    \caption{\fix{Overview of research design}}
    \label{fig:research-design}
\end{figure}

\fix{Our research is exploratory in the sense that we did not investigate any specific hypotheses. Instead, we propose and implement several automatic solutions and evaluate their performance using a quantitative evaluation study. Figure~\ref{fig:research-design} shows an overview of our research design.} 

\fix{We used two public datasets that we labeled manually. More details on the labeling process will be provided in Section~\ref{sec:labeling}. We used these datasets to train and test several classifiers: two simple solutions that served as baselines (keyword-driven and logistic regression) and four variations of transfer learning classifiers (BERT, RoBERTa, RemBERT, and ALBERT). We used these classifiers from the BERT family since they showed good performance in similar RE tasks~\cite{Fischbach2023,Henao2021,Hey2020,Sainani2020} and they are conveniently offered by ML libraries\footnote{We used the Simple Transformers library: \url{https://github.com/ThilinaRajapakse/simpletransformers}}. More details on the classifiers will be given in Section~\ref{sec:classifiers}.
With this research design, we want to answer the following research questions:}

\textbf{RQ1: What performance do automatic classifiers achieve in identifying and distinguishing Kano model factors in app reviews?}
To answer this research question, we evaluate several classifiers (simpler and BERT-based) and compare their performance. 

\textbf{RQ2: How well do the classifiers generalize when applied to unseen app reviews?}
To answer this research question, we evaluate each model on a dataset that is completely different from the training data with respect to the contained reviews and to the time period in which they were collected. Further, we perform a 10-fold cross-validation on a combination of both datasets to investigate whether a more diverse dataset improves the classifiers' performance.

\textbf{RQ3: Does misclassification of automatic classifiers correlate with disagreement of human judgement?}
To answer this research question, we assess the relationship between the cases that the classifier labeled (in)correctly and the cases where the human annotators (dis)agreed.

\subsection{Studied Datasets and Manual Labeling}
\label{sec:labeling}

For our work, we examined two independent datasets of app reviews published in the Apple App Store (iOS) and the Google Play Store (Android). The first data set (called \textit{\Stanik~}in the following) was assembled by Stanik~et~al.~\cite{Stanik19} and contains 6,070 reviews. The second data set (called \textit{\Brunotte~}in the following) was assembled by Brunotte~\cite{brunotte22}. While the recently published dataset~\cite{brunotte22} contains a lot more reviews, we worked on a subset of 1,622 of these reviews that the authors sent us earlier.

We manually labeled all reviews according to the Kano model either as ``basic'', ``performance'', ``delighter'', or ``irrelevant''. 
We considered both functional (i.e., implemented) and dysfunctional (i.e., not implemented) features. We labeled the reviews by considering only the review text.
To ensure a consistent labeling, we designed a labeling guideline that is shown in Table~\ref{table:kano-guideline}.
If a review contained indications of more than one factor, we focused on the factor that was most prominent.

\begingroup
    \renewcommand*{\arraystretch}{0.5}
    \begin{table}
    \centering
    \caption{Kano labeling guideline}
    \label{table:kano-guideline}
    \begin{tabularx}{\textwidth}{@{\extracolsep{5pt}}XX@{}} 
    \toprule
     \textbf{Basic} & \textbf{Performance}\\
    \begin{compactitem}
        \item user discontinues using the app or switches to alternative
        \item app is not usable (e.g., crashes, log in not possible)
        \item lack of a basic feature results in a bad rating
    \end{compactitem} 
     & \begin{compactitem} 
        \item moderate amount of expressed joy / annoyance
        \item constructive criticism / suggestions
    \end{compactitem}\\[1em]
     \textbf{Delighter} & \textbf{Irrelevant}\\
    \begin{compactitem}
        \item app is favored and recommended over similar apps due to these features
        \item user is a long term user due to these features
        \item praise or suggestion for addition
    \end{compactitem}
     & \begin{compactitem}
         \item cannot be labeled as any other category
         \item no clear reference to a distinct feature
     \end{compactitem}\\
    \bottomrule
    \end{tabularx}
    \end{table}
\endgroup

The reviews of the \Brunotte~have been labeled by two independent researchers. The inter-rater agreement in terms of Cohen's Kappa was $\kappa=0.7$, which may represent a ``substantial agreement''~\cite{Landis1977}. If the two labelers disagreed, we involved a third labeler as a tie breaker. 
In the labeled dataset, we marked whether the label of a review was unanimously assigned or if there was a need for a tie breaker. 
The resulting distributions of labels in the two datasets are depicted in Table~\ref{table:labels}.

\begin{table}
    \centering
    \caption{Distribution of labels in the datasets}
    \label{table:labels}
    \begin{tabular}{@{\extracolsep{3pt}}lrrrrr@{}}
    \toprule
    \textbf{Dataset}&\textbf{Total}&\textbf{Basic}&\textbf{Performance}&\textbf{Delighter}&\textbf{Irrelevant}\\
    \midrule
    \Stanik~\cite{Stanik19} & 6,070 & 1,440 & 1,530 & 648 & 2,452\\
    \Brunotte~\cite{brunotte22} & 1,622 & 1,102 & 395 & 95 & 30 \\
    \bottomrule
    \end{tabular}
\end{table}

\subsection{General Data Preprocessing} \label{preprocessing}
Data preprocessing for all classifiers consisted of the following steps:
(1) We removed duplicates, non-English reviews, and reviews that consisted only of characters but no words. (2) We converted the labels into numerical values.
Further preprocessing relevant only for specific classifiers is described in the sections of the classifiers.

\subsection{Evaluation Strategy}

To answer the research questions, we trained and tested several classifiers. For both RQ1 and RQ2, we used the \Stanik~(or a subset of it) as training set. To mitigate the class imbalance, we performed random undersampling\footnote{\fix{Random undersampling deletes examples from the majority class randomly until all classes have equally many samples.}} to create a balanced dataset with 2,592 reviews (648 reviews per class). To mitigate the random effect of undersampling, we did this five times and report the average performance metrics achieved with the five training samples. We report standard evaluation metrics (accuracy, precision, recall, $F_1$).

For RQ1 (general performance), we performed a 10-fold cross-validation on the undersampled \Stanik. 

For RQ2 (performance on unseen data), we used the undersampled \Stanik~for training and the \Brunotte~for testing. Additionally, we combined both original datasets, undersampled the combined dataset to create a balanced dataset of 2,936 reviews (743 per class), and performed a 10-fold cross-validation on this combined dataset.

For RQ3 (correlation between misclassification and human disagreement), we split the \Brunotte~into two subsets: one containing the reviews where the two labelers initially agreed on the label, and one containing the reviews where the two labelers initially disagreed on the label. We analyze the accuracy of the classifiers for these two subsets. In addition, we calculate a coefficient for the correlation between initial human disagreement and misclassifications of the classifiers. 

\section{Kano Factor Classifier}
\label{sec:classifiers}
In this section, we describe the classifiers we implemented and tested.

\subsection{Baseline Algorithms}
Besides the preprocessing described in Section~\ref{preprocessing}, we performed tokenization, stop-word removal, and computed tf-idf values for both baseline classifiers. For each non-stopword term $t$ in the training set $D$ and each label $\ell \in$ \{\texttt{basic}, \texttt{delighter}, \texttt{irrelevant}, \texttt{performance}\}, we calculated the term frequency-inverse document frequency $\texttt{tf-idf}(t,D_{\ell},D)$, where $D_\ell$ is the set of all reviews in the training set that are labeled as $\ell$.
For both approaches we used functionalities of the scikit-learn library~\cite{scikit-learn}.

\subsubsection{Keyword-driven Classifier}
We want a classifier that classifies a review based on the distribution of keywords among each class. For each review $R$ and each label $\ell$, we calculate the sum of the tf-idf values of all terms contained in the review: $M_{\ell} = \sum_{t \in R} \texttt{tfidf}(t,D_{\ell},D)$ and then categorize the review by the label $\arg\max_\ell  M_{\ell}$, which maximizes the sum of the tf-idf values.

\subsubsection{Logistic Regression}
As a second baseline algorithm, we implemented a simple tf-idf based logistic regression, where we provide the document-term matrix containing all tf-idf values per review and non-stopword as input.

\subsection{Transfer Learning Classifiers}
Transfer learning approaches use language models that have been pretrained on large sets of textual data (unsupervised learning). These language models are afterwards \textit{finetuned} with labeled data from the task and domain they are supposed to be transferred to (supervised learning). We implemented four transfer learning classifiers that use different pretrained language models namely:
\begin{compactitem}
    \item BERT~\cite{devlin-etal-2019-bert}, a language representation model for pretraining deep bidirectional representations from unlabeled text.
    \item RoBERTa~\cite{Liu2019RoBERTaAR}, an optimized version of BERT with more training data, slightly different training parameters, and different masking procedure.
    \item RemBERT~\cite{DBLP:conf/iclr/ChungFTJR21}, a pretrained language model with decoupled embeddings.
    \item ALBERT~\cite{article}, a lite variant of BERT with less parameters resulting in faster training and less memory consumption.
\end{compactitem}
By using the Simple Transformers library\footnotemark[4] to implement these classifiers, additional preprocessing was kept to a minimum, as this is already handled by the library. We used the default values for all hyperparameters.

\section{Results}
In this section, we present the results of our \fix{evaluation}. For this, we divide the section into three parts, each covering one of our research questions. We performed each run on five different undersampled sets. The values in the tables are averages of each reported metric of the five runs.

\subsection{RQ1: Performance of Classifiers}

\begin{table}
    \centering
    \renewcommand{\arraystretch}{1.1}
    \caption{Performance results from a 10-fold cross-validation on \Stanik{} in terms of accuracy (Acc.) and precision (\Precision), recall (\Recall), and $F_1$-score (\FOne) for each label.}
    \label{table:results}
    \begin{tabularx}{\textwidth}{@{\extracolsep{1pt}}Xrrrrrrrrrrrrr@{}}
    \toprule
          &  & \multicolumn{3}{c}{\textbf{Basic}} & \multicolumn{3}{c}{\textbf{Performance}} & \multicolumn{3}{c}{\textbf{Delighter}} & \multicolumn{3}{c}{\textbf{Irrelevant}} \\
         \cmidrule(lr){3-5}
         \cmidrule(lr){6-8}
         \cmidrule(lr){9-11}
         \cmidrule(lr){12-14}
         \textbf{Classifier}
         & \textbf{Acc.}
         &\Precision & \Recall & \FOne
         &\Precision & \Recall & \FOne 
         &\Precision & \Recall & \FOne 
         &\Precision & \Recall & \FOne 
         \\ 
         \midrule
         Keyword-Driven  & \fix{.514} & \fix{.589} & \fix{.793} & \fix{.675} & \fix{.494} & \fix{.521} & \fix{.505} & \fix{.411} & \fix{.552} & \fix{.467} & \fix{.810} & \fix{.194} & \fix{.303}\\
         Logistic Regression & \fix{.663} & \fix{.790} & \fix{.824} & \fix{.805} & \fix{.568} & \fix{.535} & \fix{.549} & \fix{.587} & \fix{.559} & \fix{.568} & \fix{.698} & \fix{.735} & \fix{.714}\\
         RoBERTa & .918 & .951 & .967 & .959 & .880 & .851 & .864 & .875 & \textbf{.928} & .899 & \textbf{.972} & .924 & .947 \\
         BERT & \textbf{.928} & \textbf{.960} & \textbf{.972} & \textbf{.966} & \textbf{.896} & \textbf{.871} & \textbf{\fix{.883}} & \textbf{.894} & .927 & \textbf{\fix{.910}} & .964 & \textbf{.941} & \textbf{.952}\\
         RemBERT & .633 & .586 & .673 & \fix{.626} & .518 & .550 & \fix{.534} & .538 & .630 & \fix{.580} & .596 & .683 & \fix{.637}\\
         ALBERT & .838 & .893 & .928 & .909 & .760 & .721 & \fix{.740} & .796 & .823 & \fix{.809} & .901 & .878 & .888\\
         \bottomrule
    \end{tabularx}
\end{table}

Table~\ref{table:results} shows the performance metric scores each classifier achieved in a 10-fold cross-validation on the \Stanik. 
The BERT classifier outperformed both baseline classifiers and RemBERT by a magnitude, with an accuracy of 92.8\%. BERT performs best across almost all labels and metrics, but RoBERTa's scores are very close to those of BERT. RoBERTa achieved slightly higher scores for recall in delighter features (+0.1\%) and precision in irrelevant features (+0.8\%). 
To answer RQ1, we can summarize that automated classifiers can identify and distinguish Kano model factors with an accuracy of up to 92.8\%.

\subsection{RQ2: Generalization to Unseen Data}

Table~\ref{table:results-testset} shows the performance results of all tested classifiers when trained on the entire \Stanik{} and tested on the \Brunotte. RoBERTa generally performed best with an accuracy of 72.5\%, \fix{but for irrelevant features, RemBERT achieved a better recall than any other approach and ALBERT achieved the highest precision. Also, for delighters, BERT achieved a better recall than RoBERTa}.
\begin{table}
    \centering
    \renewcommand{\arraystretch}{1.1}
    \caption{Performance results from a validation on the \Brunotte~(training on \Stanik) in terms of accuracy (Acc.) and precision (\Precision), recall (\Recall), and $F_1$-score (\FOne) for each label.}
    \label{table:results-testset}
    \begin{tabularx}{\textwidth}{@{\extracolsep{1pt}}Xrrrrrrrrrrrrr@{}}
    \toprule
          &  & \multicolumn{3}{c}{\textbf{Basic}} & \multicolumn{3}{c}{\textbf{Performance}} & \multicolumn{3}{c}{\textbf{Delighter}} & \multicolumn{3}{c}{\textbf{Irrelevant}} \\
         \cmidrule(lr){3-5}
         \cmidrule(lr){6-8}
         \cmidrule(lr){9-11}
         \cmidrule(lr){12-14}
         \textbf{Classifier}
         & \textbf{Acc.}
         &\Precision & \Recall & \FOne
         &\Precision & \Recall & \FOne 
         &\Precision & \Recall & \FOne 
         &\Precision & \Recall & \FOne 
         \\ 
         \midrule
         Keyword-Driven & \fix{.593} & \fix{.800} & \fix{.741} & \fix{.770} & \fix{.513} & \fix{.208} & \fix{.296} & \fix{.143} & \fix{.663} & \fix{.235} & \fix{.000} & \fix{.000} & \fix{.000}\\
         Logistic Regression & \fix{.600} & \fix{.858} & \fix{.689} & \fix{.764} & \fix{.451} & \fix{.360} & \fix{.400} & \fix{.181} & \fix{.726} & \fix{.289} & \fix{.075} & \fix{.100} & \fix{.086} \\
         RoBERTa & \textbf{.725} & \textbf{.896} & \textbf{.783} & \textbf{.836} & \textbf{.515} & \textbf{.599} & \textbf{.553} & \textbf{.391} & .781 & \textbf{.520} & .353 & .093 & \fix{.147}\\
         BERT & .682 & .895 & .734 & .806 & .461 & .556 & \fix{.504} & .336 & \textbf{.804} & \fix{.473} & .381 & .066 & .112\\
         RemBERT & .488 & .538 & .475 & .504 & .354 & .558 & \fix{.433} & .225 & .423 & .293 & .304 & \textbf{.233} & \textbf{\fix{.264}}\\
         ALBERT & .660 & .885 & .715 & .789 & .432 & .523 & \fix{.473} & .302 & .770 & .433 & \textbf{.466} & .040 & .072\\
         \bottomrule
    \end{tabularx}
\end{table}
Table~\ref{table:results-combined} shows the results of a 10-fold cross validation on the combination of the \Stanik{} and the \Brunotte{}. Here, BERT is the clear winner, as it performed best across all labels and metrics, with the only exception being that RoBERTa, with +0.1\%, had a negligible higher precision on irrelevant features. Again, the scores of BERT and RoBERTa are very close together. BERT achieves a very good accuracy of 95.7\%, which is a huge improvement over the \fix{logistic regression classifier as best-performing baseline, only achieving an accuracy of 60\%}.

\begin{table}
    \centering
    \renewcommand{\arraystretch}{1.1}
    \caption{Performance results from a 10-fold cross validation on the combined dataset (\Stanik{} and \Brunotte) in terms of accuracy (Acc.) and precision (\Precision), recall (\Recall), and $F_1$-score (\FOne) for each label.}
    \label{table:results-combined}
    \begin{tabularx}{\textwidth}{@{\extracolsep{1pt}}Xrrrrrrrrrrrrr@{}}
    \toprule
          &  & \multicolumn{3}{c}{\textbf{Basic}} & \multicolumn{3}{c}{\textbf{Performance}} & \multicolumn{3}{c}{\textbf{Delighter}} & \multicolumn{3}{c}{\textbf{Irrelevant}} \\
         \cmidrule(lr){3-5}
         \cmidrule(lr){6-8}
         \cmidrule(lr){9-11}
         \cmidrule(lr){12-14}
         \textbf{Classifier}
         & \textbf{Acc.}
         &\Precision & \Recall & \FOne
         &\Precision & \Recall & \FOne 
         &\Precision & \Recall & \FOne 
         &\Precision & \Recall & \FOne 
         \\ 
         \midrule
         Keyword-Driven & \fix{.482} & \fix{.475} & \fix{.826} & \fix{.602} & \fix{.436} & \fix{.466} & \fix{.445} & \fix{.452} & \fix{.428} & \fix{.438} & \fix{.822} & \fix{.207} & \fix{.330}\\
         Logistic Regression & \fix{.655} & \fix{.723} & \fix{.811} & \fix{.764} & \fix{.544} & \fix{.499} & \fix{.519} & \fix{.600} & \fix{.574} & \fix{.586} & \fix{.725} & \fix{.728} & \fix{.725} \\
         RoBERTa & .946 & .965 & .976 & .969 & .926 & .900 & .911 & .914 & .952 & .932 & \textbf{.984} & .960 & .971\\
         BERT & \textbf{.957} & \textbf{.971} & \textbf{.983} & \textbf{.978} & \textbf{.944} & \textbf{.923} & \textbf{.932} & \textbf{.940} & \textbf{.960} & \textbf{.949} & .983 & \textbf{.972} & \textbf{.977}\\
         RemBERT & .528 & .416 & .509 & \fix{.458} & .399 & .468 & \fix{.431} & .413 & .560 & \fix{.474} & .462 & .580 & \fix{.514}\\
         ALBERT & .891 & .923 & .946 & .934 & .839 & .819 & \fix{.829} & .866 & .879 & \fix{.872} & .934 & .923 & .928\\
         \bottomrule
    \end{tabularx}
\end{table}

Comparing the data from Tables~\ref{table:results-testset}~and~\ref{table:results-combined}, we can see significant differences. 
In the 10-fold cross-validation setting, performed on the combination of both datasets, we achieved very good results. In the setting where we trained on the \Stanik{} and evaluated on the \Brunotte{}, we see significantly lower metric scores. This indicates that, while the classifiers perform good when applied to unseen but ``similar'' data, this is not the case when evaluated on ``unsimilar'' data. By ``similar'', we mean the characteristics of the datasets, as each have been collected in different time periods, both cover different apps, etc. 
We specifically see problems when applied to ``unsimilar'' data in terms of delighter and irrelevant features, which is not the case in the other two evaluation settings. This may be due to the time span that lies between the collection of the \Stanik{} (2013--2015) and the \Brunotte{} (2021), as language and culture in app reviews might have evolved, but also in accordance with the Kano model, features that once were delighters now fall in different categories and once irrelevant features may have become relevant. In contrast, features, which then were basic features may still be basic features nowadays.
To answer RQ2, we can say that generalization to unseen and dissimilar app reviews is moderate. Possible solution or mitigation approaches are discussed in Section~\ref{section:discussion}.

\subsection{RQ3: Correlation between Misclassification and Human Disagreement}

Table~\ref{table:results-misclassification} shows the accuracy of the classifiers when trained on the \Stanik{} and tested on the reviews of the \Brunotte{} with consistent initial labels and inconsistent initial labels. Further, for each classifier we computed the \textit{phi} coefficient (or mean square contingency coefficient) to denote the correlation between misclassification and ambiguity of manual labeling. As binary variables, we used $\mathit{Mis} \in \{0,1\} = \{ 1 \textrm{ if classification is incorrect}, 0 \textrm{ otherwise}\}$ and $\mathit{Diff} \in \{0,1\} = \{1 \textrm{ if initial labels have been different}, 0 \textrm{ otherwise}\}$. The phi coefficient is given in the last column of Table~\ref{table:results-misclassification}. 

\begin{table}
    \centering
    \renewcommand{\arraystretch}{1.1}
    \caption{Accuracy results for classification on labels with initial agreement vs.\ initial disagreement.}
    \label{table:results-misclassification}
    \begin{tabular}{@{\extracolsep{3pt}}lrrr@{}}
    \toprule
         \textbf{Classifier} & \textbf{Accuracy} & \textbf{Accuracy} & \textbf{Phi}\\
         &\textbf{(agreed labels) }& \textbf{(disagr.\ labels)}& \textbf{Coeff.}\\
    \midrule
         Keyword-Driven  & .600 & .292 & .219 \\
         Logistic Regression & .642 & .293 & .192\\
         RoBERTa         & \textbf{.775} & \textbf{.437} & \textbf{.269} \\
         BERT & .726 & .428 & .229 \\
         RemBERT & .509 & .362 & .094 \\
         ALBERT & .703 & .405 & .224 \\
    \bottomrule
    \end{tabular}
\end{table}

We see a significant difference in the accuracy of labels where the raters agreed vs.\ labels where they disagreed, both for the baselines and for the BERT-based classifiers. For RemBERT and the logisitic regression, we observed a negligible positive  correlation ($< 0.2$).
For the keyword-based classifier, RoBERTa, BERT and ALBERT, we observed a weak positive correlation (0.2--0.3). 

To answer RQ3, we can say that all classifiers performed worse in terms of accuracy on reviews that caused initial disagreement among human annotators. The phi coefficients, though, are not impressively high and thus a strong correlation is not clearly visible.

\section{Discussion} \label{section:discussion}
\fix{In this section, we discuss our findings and its impact for research and practice. We provide a critical analysis of its strengths, weaknesses, and the threats to validity.}

\subsection{\fix{Impact in Practice}}

\fix{We think that our results are promising and show the potential for an automatic solution with sufficient performance. In comparison to the largely manual original Kano model analysis, an automatic approach is cheaper and scales better to large sets of user feedback. The results may support requirements engineering and decision making. }

\textbf{\fix{Comparison with existing approaches.}} 
\fix{Existing automated approaches~\cite{AlAmoudi2022,Lee2022} try to mimic the original Kano analysis by identifying the sentiment in user feedback, clustering it according to latent topics, and finally relating it to certain product features. The reported evaluations, however, indicate low predictive performance. AlAmoudi~et~al.~\cite{AlAmoudi2022} report $F_1$-scores between 0.30 and 0.63 for three Kano model factors. Lee~et~al.~\cite{Lee2022} did not perform a quantitative evaluation but report that ``some meaningful results are found [\ldots] This resulted in the limitation of quantification of ‘Topic Modeling’ method.''
We follow a different approach and use supervised learning to predict the Kano model factors directly from the text. Our results showed $F_1$ scores above 0.9 for all Kano model factors in the 10-fold cross validation on our combined dataset. Since the authors of the two papers did not share their datasets, we were not able to perform a direct comparison but we are confident that our approach would outperform the existing approaches also in a direct comparison. 
}

\textbf{Reviews are missing contextual information.} 
Our analysis of the labeling process and the performance of the classifiers show that neither humans nor any of our classifiers were able to always predict the Kano labels correctly (w.r.t. to what our truth set defines as correct). This may be an inherent problem of our approach to assess the Kano factors purely based on the text of a review. App reviews are limited in size, lack contextual information, and do not offer possibilities for further inquiries~\cite{Maalej2016a}. Therefore, the tone and sentiment that is conveyed with the review also plays a role for assessing the factor. Apparently, some of our tested classifiers were able to incorporate this at least to some degree. On the other hand, assessing Kano model factors by analyzing app reviews may always be less clear than the original assessment via specific interviews or surveys. 

\textbf{Application and usefulness of the approach.} 
While our study focuses on the feasibility and the performance of the classifiers, an open question is how our approach is perceived in a real world setting. We envision our approach to be used as an assistant tool for requirements analysts. Therefore, the effectiveness needs to be assessed in its context of use. Results from our performance evaluation may not be transferable directly to its context of use since the use of tools also affects working habits and perceptions of analysts (cf.~\cite{Winkler2018}). 

\subsection{\fix{Impact for Research}}

\textbf{Testing generalization on unseen data is important.}
Many studies indicate the need to test classifiers on unseen and unconsidered data. Still, this practice is not very common in software engineering research. Just recently, Dell'Anna~et~al.~\cite{DellAnna2022} showed the importance of this step by applying two recently published classifiers to unseen data and observed a significant decrease in performance. In our study, we have seen a similar degradation in performance when training on the \Stanik{} and testing on the \Brunotte{}, although the two datasets are conceptually very similar and both are already diversified by incorporating reviews of several apps. This result shows that more data may be needed to train a classifier that generalizes well.

\textbf{Hard for humans, hard for the machine.}
Our results show a correlation between misclassifications and human disagreement. This suggests that reviews that were hard to classify for humans tend to also be hard to classify for the machine. This is consistent for all classifiers that we tested. To illustrate this, consider the following review:

\noindent \textit{``Most convenient calories counter app I've ever used and I've probably used them all, super easy to add foods on your own and also cheap, super recommended''}

We labeled this review as performance factor although one of the two labelers identified it as delighter. Most classifiers also classified it as delighter. 
The reviewer is very happy about the app (high satisfaction) and favors it over other apps mainly due to easier-to-use features and cheaper price, which are classical performance factors. However, the high level of excitement may also indicate that these feature are real delighters, which the reviewer has never experienced in other apps.
Here is another example:

\noindent\textit{``Can you please update this with the map of Bhutan \& Nepal.. I am going to drive to bhutan this October but I can't find any bhutan map which could be useful to work in offline.. Please update us quickly..''} 

We labeled this review as delighter although one of the two labelers identified it as performance factor. Some classifiers classified it as performance factor while others even predicted it to be a basic feature.
The reviewer is asking for an urgently demanded but rare feature, which may indicate that this is really a delighter for the reviewer. However, the review is also phrased in a way that suggests some disappointment (``I can't find'', ``Please update us quickly..''). Also, it is not clear from the review whether other maps are available offline and, thus, adding specific regions  creates proportionately more satisfaction (i.e., performance factor).




\subsection{Threats to Validity}
Here, we discuss potential threats to validity of our models and evaluation.

\textbf{Construct validity}.
The Kano model, as defined by Kano~\cite{kano1984} consists of five classes: delighter, performance, basic, irrelevant, and rejection features. In this paper, we did not consider the rejection feature class, since they are fairly rare and their definition is ambiguous. Still, there may be reviews that fall into this class. Examples include features that are annoying to users, e.g., excessive advertisement, but also features that users perceive as threat to their privacy, e.g., app tracking or the ``blue read-checkmark'' in messenger apps.

\textbf{Internal validity}.
In our manual labeling process, we assigned each review exactly one label. This, however, might be too coarse grained, as some reviews contain more than one Kano factor, e.g., a user reports a problem with a basic feature but also suggests ideas that are delighters. 
Reviews containing more than one factor are problematic, as they can be classified ambiguously by the classifiers and human annotators. There are two main approaches to solve this problem: (1) \textit{Separation:} We separate each aspect of a complex review as a distinct review. This, however, may break contextual links between aspects. Also, the boundary between them is often blurred. (2) \textit{Multilabel classification:} When a review covers multiple Kano factors, we can assign multiple labels accordingly. This introduces a huge overhead and thus is only feasible when a significant number of reviews include multiple different factors. Usually, this is not the case.
In our labeling process, we decided to assign labels according to the most prominent aspect of a review, e.g., if a user reports a bug that makes the app unusable and expresses their dissatisfaction, but also reports a delighter, we assigned the basic label.

Despite our efforts to make the labeling process as transparent and systematic as possible, there may still be some variability in the resulting gold standard, e.g., misinterpretation of the users intention, blurred boundaries of the Kano factors, too broad or too narrow judgement or human mistakes.

\textbf{External validity}.
Our results have shown that generalization of our tested classifiers is fairly moderate when applied to unseen, dissimilar test data. This may indicate that more data is needed to train a classifier that generalizes better. 

Lastly, app reviews are not the only relevant source of user feedback.
\newline Nayebi~et~al.~\cite{Nayebi2017} mined 70 apps for six weeks on app store reviews and on Twitter. They found that Twitter provided 22.4\% more feature requests and 12.9\% more bug reports.

\section{Conclusions}
In this paper, we presented an automated approach to app review classification according to the Kano model. We evaluated several BERT-based models and found that, overall, BERT performed best with an accuracy of 92.8\% to 95.7\%. We compared our findings to two baseline approaches based on traditional machine learning techniques and found that most BERT-based classifier outperform them by magnitudes, i.e., by \fix{around 30}\%. We evaluated the generalization of our classifier to unseen app reviews and found that the performance of all classifiers dropped significantly. We conclude that more data is needed to achieve a classifier that performs well on unseen data. We also evaluated that misclassification of the classifiers does, to some degree, correlate with ambiguity in the manual labeling process, as accuracy differs by 33.8\% between consistent and inconsistent labeling.
What is still missing in our work is an evaluation of the approach in its context, i.e., in terms of a user study.

\subsubsection{Acknowledgements} We want to thank the authors of the two datasets for permission to use parts of their dataset and the permission to publish our labeled dataset. We also want to thank Murat Sancak for his initial work on the topic in his Bachelor's thesis.

%
%
%
\bibliographystyle{splncs04}
\bibliography{references}

\begin{thebibliography}{10}
\providecommand{\url}[1]{\texttt{#1}}
\providecommand{\urlprefix}{URL }
\providecommand{\doi}[1]{https://doi.org/#1}

\bibitem{Achimugu2014}
Achimugu, P., Selamat, A., Ibrahim, R., Mahrin, M.N.: A systematic literature
  review of software requirements prioritization research. Information and
  Software Technology  \textbf{56}(6),  568--585 (2014).
  \doi{10.1016/j.infsof.2014.02.001}

\bibitem{AlAmoudi2022}
AlAmoudi, N., Baslyman, M., Ahmed, M.: Extracting attractive app aspects from
  app reviews using clustering techniques based on kano model. In: {IEEE}
  International Requirements Engineering Conference Workshops ({REW}). {IEEE}
  (2022). \doi{10.1109/REW56159.2022.00030}

\bibitem{brunotte22}
Brunotte, W.: App store reviews (Nov 2022). \doi{10.5281/zenodo.7319510}

\bibitem{Bukhsh2020}
Bukhsh, F.A., Bukhsh, Z.A., Daneva, M.: A systematic literature review on
  requirement prioritization techniques and their empirical evaluation.
  Computer Standards {\&} Interfaces  \textbf{69} (2020).
  \doi{10.1016/j.csi.2019.103389}

\bibitem{DBLP:conf/iclr/ChungFTJR21}
Chung, H.W., F{\'{e}}vry, T., Tsai, H., Johnson, M., Ruder, S.: Rethinking
  embedding coupling in pre-trained language models. In: 9th International
  Conference on Learning Representations (ICLR). OpenReview.net (2021)

\bibitem{Dalpiaz2018}
Dalpiaz, F., Ferrari, A., Franch, X., Palomares, C.: Natural language
  processing for requirements engineering: The best is yet to come. {IEEE}
  Software  \textbf{35}(5),  115--119 (2018). \doi{10.1109/ms.2018.3571242}

\bibitem{DellAnna2022}
Dell'Anna, D., Aydemir, F.B., Dalpiaz, F.: Evaluating classifiers in {SE}
  research: the {ECSER} pipeline and two replication studies. Empirical
  Software Engineering  \textbf{28}(1) (2022). \doi{10.1007/s10664-022-10243-1}

\bibitem{devlin-etal-2019-bert}
Devlin, J., Chang, M.W., Lee, K., Toutanova, K.: {BERT}: Pre-training of deep
  bidirectional transformers for language understanding. In: Proceedings of the
  2019 Conference of the North {A}merican Chapter of the Association for
  Computational Linguistics: Human Language Technologies, Volume 1 (Long and
  Short Papers). pp. 4171--4186. Association for Computational Linguistics
  (2019). \doi{10.18653/v1/N19-1423}

\bibitem{Fischbach2023}
Fischbach, J., Frattini, J., Vogelsang, A., Mendez, D., Unterkalmsteiner, M.,
  Wehrle, A., Henao, P.R., Yousefi, P., Juricic, T., Radduenz, J., Wiecher, C.:
  Automatic creation of acceptance tests by extracting conditionals from
  requirements: {NLP} approach and case study. Journal of Systems and Software
  \textbf{197} (2023). \doi{10.1016/j.jss.2022.111549}

\bibitem{Groen2017}
Groen, E.C., Seyff, N., Ali, R., Dalpiaz, F., Doerr, J., Guzman, E., Hosseini,
  M., Marco, J., Oriol, M., Perini, A., Stade, M.: The crowd in requirements
  engineering: The landscape and challenges. {IEEE} Software  \textbf{34}(2),
  44--52 (2017). \doi{10.1109/ms.2017.33}

\bibitem{Guzman2014}
Guzman, E., Maalej, W.: How do users like this feature? a fine grained
  sentiment analysis of app reviews. In: {IEEE} 22nd International Requirements
  Engineering Conference ({RE}) (2014). \doi{10.1109/re.2014.6912257}

\bibitem{Henao2021}
Henao, P.R., Fischbach, J., Spies, D., Frattini, J., Vogelsang, A.: Transfer
  learning for mining feature requests and bug reports from tweets and app
  store reviews. In: {IEEE} International Requirements Engineering Conference
  Workshops ({REW}). {IEEE} (2021). \doi{10.1109/rew53955.2021.00019}

\bibitem{Herrmann2008}
Herrmann, A., Daneva, M.: Requirements prioritization based on benefit and cost
  prediction: An agenda for future research. In: {IEEE} International
  Requirements Engineering Conference (RE) (2008). \doi{10.1109/re.2008.48}

\bibitem{herzberg1993motivation}
Herzberg, F., Mausner, B., Snyderman, B.: The motivation to work. Transaction
  Pub (1993)

\bibitem{Hey2020}
Hey, T., Keim, J., Koziolek, A., Tichy, W.F.: {NoRBERT}: Transfer learning for
  requirements classification. In: {IEEE} International Requirements
  Engineering Conference ({RE}) (2020). \doi{10.1109/re48521.2020.00028}

\bibitem{Hujainah2018}
Hujainah, F., Bakar, R.B.A., Abdulgabber, M.A., Zamli, K.Z.: Software
  requirements prioritisation: A systematic literature review on significance,
  stakeholders, techniques and challenges. {IEEE} Access  \textbf{6},
  71497--71523 (2018). \doi{10.1109/access.2018.2881755}

\bibitem{kano1984}
Kano, N., Seraku, N., Takahashi, F., Tsuji, S.: Attractive quality and must-be
  quality. Journal of the Japanese Society for Quality Control  \textbf{14}(2),
   147--156 (1984)

\bibitem{article}
Lan, Z., Chen, M., Goodman, S., Gimpel, K., Sharma, P., Soricut, R., Research,
  G., de~Carvalho, M.: Albert: A lite bert for self-supervised learning of
  language representations  (10 2019)

\bibitem{Landis1977}
Landis, J.R., Koch, G.G.: The measurement of observer agreement for categorical
  data. Biometrics  \textbf{33}(1) (1977). \doi{10.2307/2529310}

\bibitem{Lee2022}
Lee, H., Cha, M.S., Kim, T.: Text mining-based mapping for kano quality factor.
  ICIC Express Letters. Part B, Applications: an International Journal of
  Research and Surveys  \textbf{12}(2),  185--191 (2021)

\bibitem{Lim2021SLR}
Lim, S., Henriksson, A., Zdravkovic, J.: Data-driven requirements elicitation:
  A systematic literature review. SN Computer Science  \textbf{2:16} (2021)

\bibitem{Liu2019RoBERTaAR}
Liu, Y., Ott, M., Goyal, N., Du, J., Joshi, M., Chen, D., Levy, O., Lewis, M.,
  Zettlemoyer, L., Stoyanov, V.: Roberta: A robustly optimized bert pretraining
  approach. ArXiv  \textbf{abs/1907.11692} (2019)

\bibitem{Maalej2016b}
Maalej, W., Kurtanovi{\'{c}}, Z., Nabil, H., Stanik, C.: On the automatic
  classification of app reviews. Requirements Engineering  \textbf{21}(3),
  311--331 (2016). \doi{10.1007/s00766-016-0251-9}

\bibitem{Maalej2015}
Maalej, W., Nabil, H.: Bug report, feature request, or simply praise? on
  automatically classifying app reviews. In: {IEEE} 23rd International
  Requirements Engineering Conference ({RE}) (2015).
  \doi{10.1109/re.2015.7320414}

\bibitem{Maalej2016a}
Maalej, W., Nayebi, M., Johann, T., Ruhe, G.: Toward data-driven requirements
  engineering. {IEEE} Software  \textbf{33}(1),  48--54 (2016).
  \doi{10.1109/ms.2015.153}

\bibitem{Nayebi2017}
Nayebi, M., Cho, H., Farrahi, H., Ruhe, G.: App store mining is not enough. In:
  {IEEE}/{ACM} International Conference on Software Engineering Companion
  ({ICSE}-C) (2017). \doi{10.1109/icse-c.2017.77}

\bibitem{Pagano2013}
Pagano, D., Maalej, W.: User feedback in the appstore: An empirical study. In:
  {IEEE} International Requirements Engineering Conference ({RE}) (2013).
  \doi{10.1109/re.2013.6636712}

\bibitem{scikit-learn}
Pedregosa, F., Varoquaux, G., Gramfort, A., Michel, V., Thirion, B., Grisel,
  O., Blondel, M., Prettenhofer, P., Weiss, R., Dubourg, V., Vanderplas, J.,
  Passos, A., Cournapeau, D., Brucher, M., Perrot, M., Duchesnay, E.:
  Scikit-learn: Machine learning in {P}ython. Journal of Machine Learning
  Research  \textbf{12},  2825--2830 (2011)

\bibitem{Sainani2020}
Sainani, A., Anish, P.R., Joshi, V., Ghaisas, S.: Extracting and classifying
  requirements from software engineering contracts. In: {IEEE} International
  Requirements Engineering Conference ({RE}). {IEEE} (2020).
  \doi{10.1109/re48521.2020.00026}

\bibitem{Stanik19}
Stanik, C., Haering, M., Maalej, W.: Classifying multilingual user feedback
  using traditional machine learning and deep learning. In: IEEE International
  Requirements Engineering Conference Workshops (REW). pp. 220--226 (2019).
  \doi{10.1109/REW.2019.00046}

\bibitem{Wang2019}
Wang, C., Daneva, M., van Sinderen, M., Liang, P.: A systematic mapping study
  on crowdsourced requirements engineering using user feedback. Journal of
  Software: Evolution and Process  \textbf{31}(10),  e2199 (2019).
  \doi{https://doi.org/10.1002/smr.2199}

\bibitem{Winkler2016}
Winkler, J., Vogelsang, A.: Automatic classification of requirements based on
  convolutional neural networks. In: {IEEE} International Requirements
  Engineering Conference Workshops ({REW}) (2016). \doi{10.1109/rew.2016.021}

\bibitem{Winkler2018}
Winkler, J.P., Vogelsang, A.: Using tools to assist identification of
  non-requirements in requirements specifications {\textendash} a controlled
  experiment. In: Requirements Engineering: Foundation for Software Quality
  (REFSQ), pp. 57--71. Springer International Publishing (2018).
  \doi{10.1007/978-3-319-77243-1_4}

\bibitem{Wouters2022}
Wouters, J., Menkveld, A., Brinkkemper, S., Dalpiaz, F.: Crowd‑based
  requirements elicitation via pull feedback: method and case studies. In:
  {Requirements Engineering}. {Requirements Engineering} (2022).
  \doi{https://doi.org/10.1007/s00766-022-00384-6}

\end{thebibliography}

\end{document}